\documentclass[useAMS,usenatbib]{mn2e}
\pdfoutput=1

\usepackage{changepage}
\usepackage{booktabs}
\usepackage{aas_macros}
\usepackage{epsfig}
\usepackage{amsmath}
\usepackage{amssymb}
\usepackage{comment}
\usepackage{caption}
\usepackage[justification=centering]{caption}

\usepackage{leftidx}
\usepackage{natbib}
\usepackage[rgb]{xcolor}
\definecolor{MyGreen}{rgb}{0.0,0.6,0.3}
\definecolor{MyPurple}{rgb}{0.6,0,0.3}
\usepackage{fancyref}
\usepackage{hyperref}
\hypersetup{colorlinks=true,citecolor=MyGreen,linkcolor=MyPurple,urlcolor=blue}

\def\beq{\begin{equation}}
\def\eeq{\end{equation}}
\def\ba{\begin{eqnarray}}
\def\ea{\end{eqnarray}}
\def\bal{\begin{align}}
\def\eal{\end{align}}
\def\bxi{{\mbox{\boldmath $\xi$}}}

\begin{document}

\title[White dwarf kicks] {White Dwarf Kicks via Episodic Mass Ejection from Red Giant Stars}

\author[J. Fuller]{
Jim Fuller$^{1}$\thanks{Email: jfuller@caltech.edu} 
\\$^1$TAPIR, Mailcode 350-17, California Institute of Technology, Pasadena, CA 91125, USA
}

\label{firstpage}
\maketitle
\begin{abstract}

Recent observations have found evidence that white dwarf (WD) stars receive a kick of $\sim$1 km/s as the envelopes of their red giant progenitors are expelled. We show that these kicks can arise from asymmetric and episodic mass loss of red giant stars. Based on simple hydrodynamic models of red giant mass loss, each mass ejection event likely expels of order $\sim \! 10^{-4} \, M_\odot$, imparting a small and randomly oriented kick to the star, with the net kick to the WD arising from a combination of many $(N \sim 10^4)$ events. Each kick is of order $v_k \sim 5 \, {\rm m/s}$, with a total kick of order $v_{\rm k,tot} \approx \sqrt{N} v_k \sim  0.5 \, {\rm km/s}$. We predict substantially larger net kicks for higher mass WDs. We also model the orbital evolution of binary stars experiencing a stochastic series of kicks, developing analytic models to explain numerical integrations. For nominal mass ejection parameters, widely separated binaries $(a \gtrsim 10^3 \, {\rm AU})$ can become unbound, with larger fractions of unbound systems at larger separations and for higher WD masses. The orbital eccentricities can also approach unity, causing stellar collisions that may result in luminous transients and eccentric common envelope events.

\end{abstract}

\section{Introduction}

The vast majority of stars end their lives by expelling their hydrogen envelopes and contracting into dense and degenerate white dwarf (WD) remnants. Most stars' envelopes are predominantly lost via winds when they expand into asymptotic giant branch (AGB) stars during helium shell burning. The combination of large stellar luminosities and small surface gravities help expel AGB star envelopes, but the details of the mass loss mechanism are not fully understood.

Modern models of AGB star mass loss predict that it does not occur via a smooth and continuous wind, but instead is episodic and time-variable (see \citealt{hofner:18} and \citealt{decin:21} for reviews). A combination of coherent stellar pulsations and stochastic convection drives shock waves into the atmosphere of the star, supporting a dense, variable, and asymmetric ``chromosphere" of material above the photosphere. At distances of $\sim$3 stellar radii, the gas can cool and form dust, whose high opacity enables the star's radiation to propel a dust-driven wind. This basic mechanism has been seen repeatedly in simulations of AGB stars \citep{freytag:17,freytag:23} and red supergiants \citep{ma:25}. However, the average mass loss rate, its temporal variability, and its dependence on stellar properties (mass, radius, luminosity, pulsation period, etc.) are not fully understood. 

Interferometric observations also indicate that the stellar chromospheres are highly asymmetric for both red supergiants \citep{ohnaka:16,ohnaka:17b} and AGB stars \citep{ohnaka:25}. However, the outflows at large distances of $D \gtrsim 10^4$ AU are approximately spherically symmetric \citep{castro-carrizo:10}. It appears that individual mass ejection events are asymmetric, but averaging over many discrete events forms a quasi-symmetric outflow at large distances.

\begin{figure}
\includegraphics[scale=0.33]{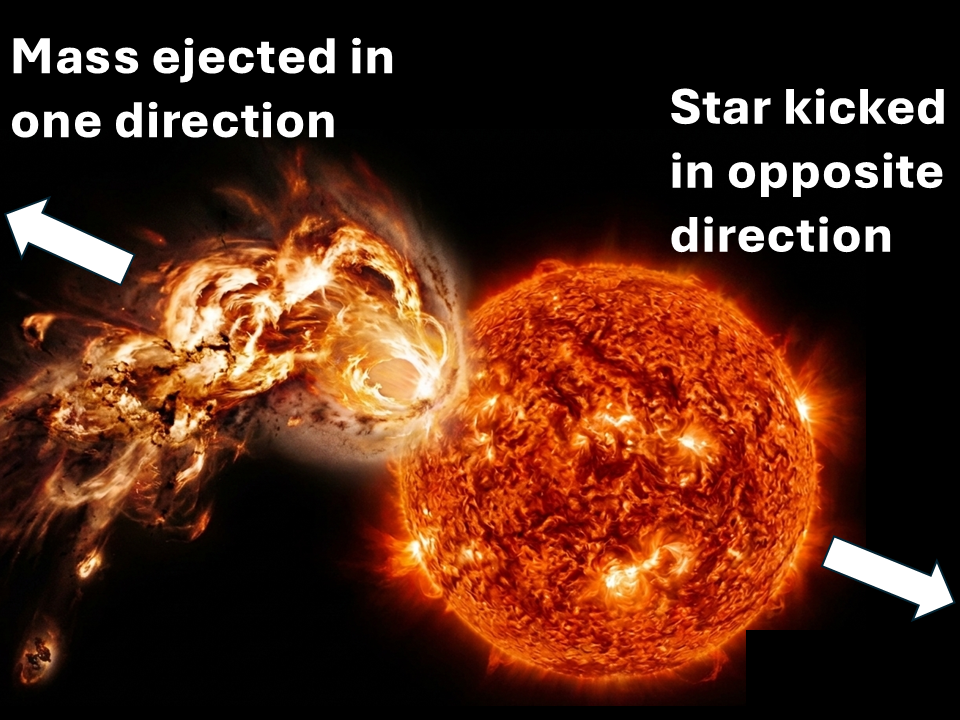}
\caption{ \label{fig:Cartoon} Cartoon showing an asymmetric mass ejection event and resulting recoil imparted to the AGB star. The combination of many randomly oriented events provides a net kick to the resulting white dwarf star.}
\end{figure}

Episodic mass loss events likely eject mass in different directions (Figure \ref{fig:Cartoon}), such that each episode imparts a small kick to the red giant star. The combination of many such episodes will produce a larger net kick to its WD descendant. These kicks may unbind long-period WD binaries \citep{elbadry:18,hwang:25} and triple systems \citep{shariat:23}. They can also eject stars from open clusters \citep{weidemann:77,fellhauer:03} and expand the spatial extent of WDs in globular clusters \citep{davis:08,heyl:07a}. 

While some of this previous work has modeled dynamical effects of WD kicks, those studies usually assumed a single, instantaneous kick to the WD at its birth, with a parameterized amplitude. One exception is \cite{caputo:26}, who considered stochastic perturbations from external bodies, and \cite{spruit:98} who considered the effects of mass ejection events on the angular momentum of an AGB star and the resulting WD. Stochastic kicks have also been considered for neutron stars, arising from asymmetries in neutrino-driven explosions \citep{scheck:06,burrows:23}, or via jittering jets that impart a series of random kicks to the neutron star \citep{shishkin:25}. 

In this work, we use simple hydrodynamic models of stellar chromospheres to estimate typical masses and kicks imparted to AGB stars by discrete mass ejection events. We compute the net kicks imparted by a series of smaller kicks, and we calculate the resulting orbital evolution for binary systems. We estimate that the net kick has an amplitude of $\sim$1 km/s with a roughly Maxwellian distribution, and that the kick amplitudes increase strongly with WD mass. These kicks can unbind widely separated binary stars, and they can eject WDs from clusters. We show that the kicks may also induce very high eccentricities and stellar collisions, leading to more dramatic events such as stellar mergers or common envelope events that produce luminous red novae.

%Early models accounted for shocks driven by stellar pulsations in 1D spherically symmetric models (e.g., \citealt{willson:79,hill:79,bowen:88}). These 1D models have been updated with more advanced treatments of radiative transfer and dust formation (e.g., \citealt{bladh:15,hofner:16,bladh:19}). The basic idea is that stellar pulsations steepen into a train of shock waves that propagate out into the star's chromosphere. The high-temperature gas behind the shocks quickly radiates to return to a near-equilibrium temperature, so it is the momentum of the shocks (rather than pressure from heated gas) that helps lift material high enough to form dust.

%In these dust-driven wind models, the mass loss rate is determined primarily by the stellar pulsations, while wind outflow speeds are affected by dust formation physics and stellar metallicity, with higher metallicity producing more dust and higher wind speeds. The models (e.g., \citealt{bladh:19}) can produce a steep increase of mass loss rate with increasing stellar luminosity $L$ and decreasing mass $M$. They also reproduce an observed correlation between mass loss rate and wind speed due to time-dependent dust grain growth in the outflow. 

%However, there are also many limitations to these models. First, the pulsations are injected by hand, and their amplitude is a free parameter which is crucial for determining the mass loss rate. Second, their 1D nature makes them incapable of accounting for multi-dimensional processes which are important for producing episodic and asymmetric mass loss as observed.

\section{Episodic Mass Ejection}
\label{sec:mass}

To estimate the kick to a red giant star imparted by a mass ejection event, we utilize the framework of \cite{fuller:24}. They built a simple model for the structure of a dense chromosphere supported by outgoing shock waves, which ejects mass episodically and asymmetrically due to the stochastic nature of the star's convection. In their model, the density profile above the surface of the star due to shocks with speed $v_s$ is 
\begin{equation}
    \rho = \rho(R) \left(\frac{R}{r}\right)^2 e^{-\frac{v_{\rm esc}^2}{2 v_{\rm s}^2} \big(1-R/r \big)} \, .
\end{equation}
Here, $\rho(R)$ is the density at the photosphere with radius $R$, and $v_{\rm esc} = \sqrt{2 G M/R}$ is the star's escape speed.

Mass ejection events occur when material is lifted up to the dust formation radius, which typically occurs at $R_{\rm d} \sim 2-4 R$ for AGB stars. These events are dominated by times when $v_s \sim \sqrt{v_{\rm esc} (1-R/R_{\rm d})^{1/2} v_{\rm con}}$, where $v_{\rm con}$ is the typical convective velocity near the photosphere. During these events, the typical density at the dust formation radius is
\begin{equation}
    \rho = \rho(R) \left(\frac{R}{R_{\rm d}}\right)^2 e^{-\frac{v_{\rm esc}}{2 v_{\rm con}}\sqrt{1-R/R_{\rm d}}} \, \, .
\end{equation}
Simulations (\citealt{freytag:17,freytag:23,ma:25}) indicate ejected plumes have a solid angle of order unity, so we expect a typical mass ejection event to carry away
\begin{align}
\label{eq:Mej}
    M_{\rm ej} &= f_{\rm ej} \, R_{\rm d}^3 \, \rho(R_d) \nonumber \\
    &= f_{\rm ej} \, \rho(R) R_{\rm d} R^2 e^{-\frac{v_{\rm esc}}{2 v_{\rm con}}\sqrt{1-R/R_{\rm d}}} \, .
\end{align}
Here, $f_{\rm ej}$ is a parameter of order unity that can be calibrated with observations of WD kicks. Plugging in numbers typical of AGB stars, $\rho(R) = 10^{-9} \, {\rm g}/{\rm cm}^3$, $M = 1.5 M_\odot$, $R = 250 \, R_\odot$, $v_{\rm con} \approx 5 \, {\rm km}/{\rm s}$, $v_{\rm esc} \approx 50 \, {\rm km}/{\rm s}$, we find $M_{\rm ej} \sim 10^{-4} \, M_\odot$. More luminous AGB stars have smaller ratios of $v_{\rm esc}/v_{\rm con}$ and therefore larger values of $M_{\rm ej}$. The actual value of $M_{\rm ej}$ (and its distribution) is uncertain at the order of magnitude level.

In order for mass to be ejected, it must be lifted to a height of $\sim \! R_{\rm d}$ above the AGB star surface, which means it must be propelled to speeds approaching $\sim \! v_{\rm esc}$. The mass can be subsequently accelerated by the radiative force on dust grains, but this will not impart a momentum kick to the star. Therefore the net momentum lost in an episodic mass ejection is $P_{\rm ej} \sim M_{\rm ej} v_{\rm esc}$, and the star must receive an equal momentum boost in the opposite direction. The kick imparted to the star thus has magnitude
\begin{align}
\label{eq:vk}
    v_{\rm k} &\sim \frac{M_{\rm ej}}{M_1} v_{\rm esc} \nonumber \\
    &\sim 3.3 \, {\rm m}/{\rm s} \, \left(\frac{M_{\rm ej}}{10^{-4} M_\odot} \right) \left(\frac{M_1}{1.5 \, M_\odot} \right)^{-1/2} \left(\frac{R}{250 R_\odot}\right)^{-1/2} \, .
\end{align}
So, individual kicks to the AGB star only have expected magnitudes of a few m/s. 

Although individual kicks are small, we expect a large number of kicks to occur, roughly $N = M_{\rm env}/M_{\rm ej} \sim 10^4$ for an ejected envelope mass $M_{\rm env} \sim 1 \, M_\odot$. We assume the kicks are uncorrelated and in random directions, so the star will undergo a random walk in each direction. For a one-dimensional random walk, the expectation value for a series of $N$ randomly signed steps scales as $\sqrt{N}$. Therefore, after $N$ mass ejection events, we expect a net kick of magnitude
\begin{align}
\label{eq:vktot}
    v_{\rm k, tot} &\sim \sqrt{N} \frac{M_{\rm ej}}{M_1} v_{\rm esc} \nonumber \\
    &\sim 0.4 \, {\rm km}/{\rm s} \, \left(\frac{M_{\rm ej}}{10^{-4} M_\odot} \right)^{\! 1/2} \! \! \left(\frac{R}{250 R_\odot}\right)^{\! -1/2} \, 
\end{align}
when the envelope mass $M_{\rm env}$ is comparable to the star's total mass $M_1$. Although these kicks are very modest compared to neutron star kicks, they are large enough to have a major impact on binary systems whose orbital velocity is $v_{\rm orb} \lesssim v_{\rm k, tot}$.

In a real star, the properties determining the kick amplitude (e.g., mass, radius, etc.) evolve with time. To estimate the net kick for an evolving star, we can reconfigure equation \ref{eq:vktot},
\begin{equation}
    \frac{d v_{\rm k,tot}^2}{dN} = v_{\rm k}^2 \, ,
\end{equation}
or in terms of mass change, 
\begin{equation}
    \frac{d v_{\rm k,tot}^2}{dM} = \frac{v_{\rm k}^2}{M_{\rm ej}} \, .
\end{equation}
Hence we can integrate over the mass lost by the star to determine the net kick, 
\begin{align}
\label{eq:vktotint}
    v_{\rm k,tot} &= \left[ \int^{M_{\rm env}}_{0} \frac{v_{\rm k}^2}{M_{\rm ej}} dM \right]^{1/2} \nonumber \\ 
    &= \left[ \int^{M_{\rm env}}_{0} \frac{M_{\rm ej}}{M_1} \frac{v_{\rm esc}^2}{M_1} dM \right]^{1/2}  \, . 
\end{align}
Here, the value of $M_{\rm ej}$ changes according to equation \ref{eq:Mej}.

\begin{figure}
\includegraphics[scale=0.37]{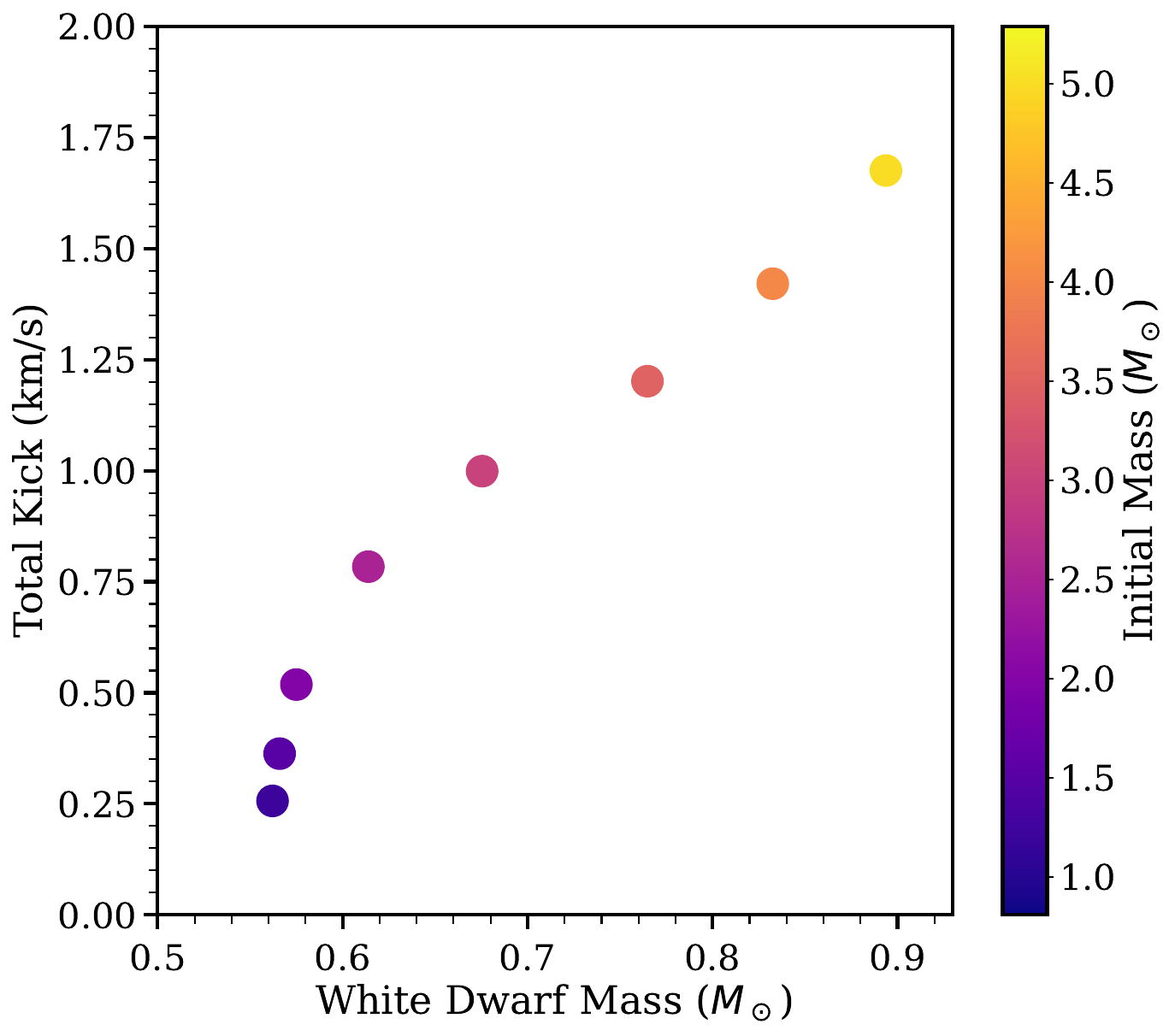}
\caption{ \label{fig:WDKicks} Estimates for net white dwarf kicks as a function of white dwarf mass. Colors indicate the initial stellar mass for each model.}
\end{figure}

To compute the expected net kicks, we generate stellar models with MESA \citep{paxton:13}, with various initial masses. We evolve them from the main sequence all the way to WDs, with the AGB mass loss rate from dust-driven winds computed as described in \cite{fuller:24}. We then compute the integral of equation \ref{eq:vktotint} from the output history files.

Figure \ref{fig:WDKicks} shows the estimated total kick imparted to WDs as a function of WD mass, using $f_{\rm ej} = 0.5$ (equation \ref{eq:Mej}). Typical values are of order $0.5 \, {\rm km/s}$, as expected from equation \ref{eq:vktot}. This is intriguingly close to the kick magnitude inferred from the distribution of WDs in wide binaries \citep{elbadry:18}. However, we caution that the net kick amplitude scales as $M_{\rm ej}^{1/2}$, which is quite uncertain.

Regardless of the value of $M_{\rm ej}$, we predict substantially larger kicks for high-mass WDs. The main reason is that the predicted values of $M_{\rm ej}$ can be larger by more than an order of magnitude for high-mass AGB stars due to their smaller values of $v_{\rm esc}/v_{\rm con}$ at the tip of the AGB. Additionally, more massive WDs are born from more massive stars that lose a larger fraction of their mass, allowing for larger kicks.

\section{Orbital Evolution Calculations}
\label{sec:orb}

\subsection{Effect of individual kicks}

Here we calculate the effect of mass ejection events on a binary's orbit. Several papers have computed the effect of kicks and mass ejection, but many are limited to circular pre-ejection orbits. We find \cite{hills:83} to be useful for the case of eccentric orbits. The primary star has mass $M_1$ and radius $R_1$, the companion star has mass $M_2$, and the semi-major axis is $a$. Each event ejects mass $M_{\rm ej}$ and imparts a kick velocity $\vec{v}_{\rm k}$ to the primary star, which we assume to occur instantaneously. 

It is useful to adopt a frame centered on and co-moving with $M_2$. In this frame, $M_1$ is located at $\vec{r}$ with initial velocity $\vec{v}$. The orbital axis is in the $z$-direction, such that the pre-kick velocity can be expressed $\vec{v} = (v_r, v_\phi, v_z)$, where $v_r$ is the radial velocity, $v_\phi$ is the tangential velocity, and $v_z=0$. Similarly, the kick velocity is $\vec{v}_{\rm k} = (v_{kr}, v_{k \phi}, v_{kz})$.

The post-kick mass of the primary star is $M_1' = M_1 - M_{\rm ej}$ and the post-kick velocity is $\vec{v}' = \vec{v} + \vec{v}_{\rm k}$. The pre-kick orbital angular momentum is $\vec{J} = M_1 M_2 \vec{r} \times \vec{v}/(M_1+M_2)$, and the post-kick angular momentum is
\begin{align}
\label{eq:jf}
    \vec{J}' &= \frac{M_1' M_2}{M_1' + M_2} \vec{r} \times \vec{v}' \nonumber \\
    &= \frac{M_1' M_2}{M_1' + M_2} r \left( 0, - v_{kz} , v_\phi + v_{k \phi} \right) \, .
\end{align}
Therefore the magnitude of the angular momentum is $J' = M_1' M_2 r \sqrt{ v_{kz}^2 + (v_\phi + v_{k \phi})^2 }/(M_1' + M_2)$.

The pre-kick orbital energy is $E = - G M_1 M_2/(2 a)$. The relative velocity between the stars at the moment of the event is
\begin{equation}
    v^2 = v_c^2 \left( \frac{2 a}{r} - 1 \right) \, ,
\end{equation}
where we have defined the circular orbital velocity 
\begin{equation}
    v_c^2 = \frac{G(M_1+M_2)}{a} \, .
\end{equation}
Following \cite{hills:83}, the orbital energy after the event is
\begin{align}
\label{eq:ef}
    E' &= - \frac{G M_1' M_2}{r} + \frac{1}{2} \frac{M_1' M_2}{M_1'+M_2} v'^2 \nonumber \\
    &= - \frac{G M_1' M_2}{2 a} \left[ \frac{2 a}{r} - \frac{M_1+M_2}{M_1'+M_2} \frac{v'^2}{v_c^2} \right] \, .
\end{align}
With the energy and angular momentum known, the semi-major axis and eccentricity at any point in time can be calculated via
\begin{align}
    a = - \frac{G M_1 M_2}{2 E} \, ,
\end{align}
\begin{align}
    e^2 = 1 + \frac{2 J^2 E (M_1 + M_2) }{ G^2 (M_1 M_2)^3} \, .
\end{align}

It is useful to compute the change in orbital energy $\Delta E$ and angular momentum $\Delta J$ due to each kick. We take the limit $M_{\rm ej} \ll M_1$ and $v_{\rm k} \ll v_c$. Then to first order in these quantities, we have
\begin{align}
\label{eq:deltae}
    \frac{\Delta E}{E} &\simeq - \frac{M_{\rm ej}}{M_1} \left[ 1 + \frac{M_1}{M_1 + M_2} \left( \frac{2 a}{r} - 1 \right) \right] \nonumber \\
    & -2 \frac{v \, v_k}{v_c^2} \cos \theta_k  - \frac{v_k^2}{v_c^2} \, .
\end{align}
Here, $\theta_k$ is the angle between the kick and the orbital velocity of star 1. The last term in equation \ref{eq:deltae} is second-order in $v_k$. It must be retained because its sign is always negative, whereas first-order terms in $v_k$ have randomly varying signs. Hence, that term will be important for long-term evolution as shown in Section \ref{sec:analytical}.

To first order, the change in angular momentum is
\begin{equation}
\label{eq:deltaj}
    \frac{\Delta \vec{J}}{J} \simeq - \frac{M_{\rm ej}}{M_1} \frac{M_2}{M_1 + M_2} \hat{z} + \frac{v_{k \phi}}{v_\phi} \hat{z} - \frac{v_{kz}}{v_\phi} \hat{\phi} \, .
\end{equation}
Note the first two terms affect the magnitude of the angular momentum, which may increase or decrease. The last term affects primarily the direction of the angular momentum, causing the orbital axis to change direction.

\begin{figure}
\includegraphics[scale=0.36]{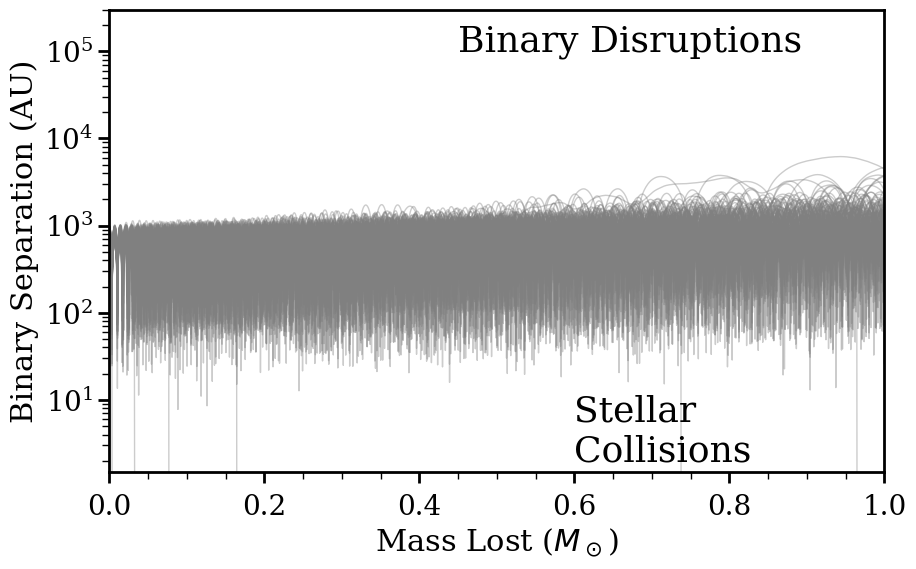}
\includegraphics[scale=0.36]{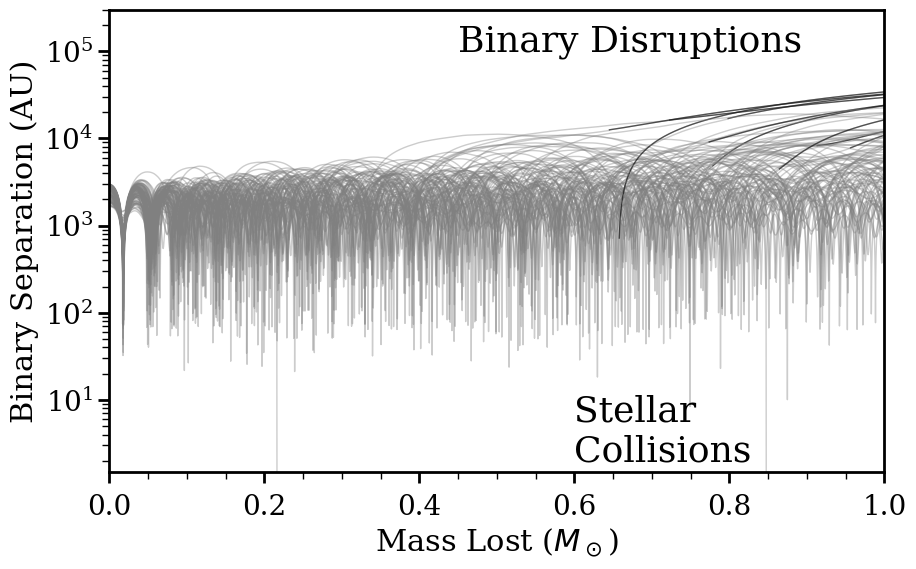}
\includegraphics[scale=0.36]{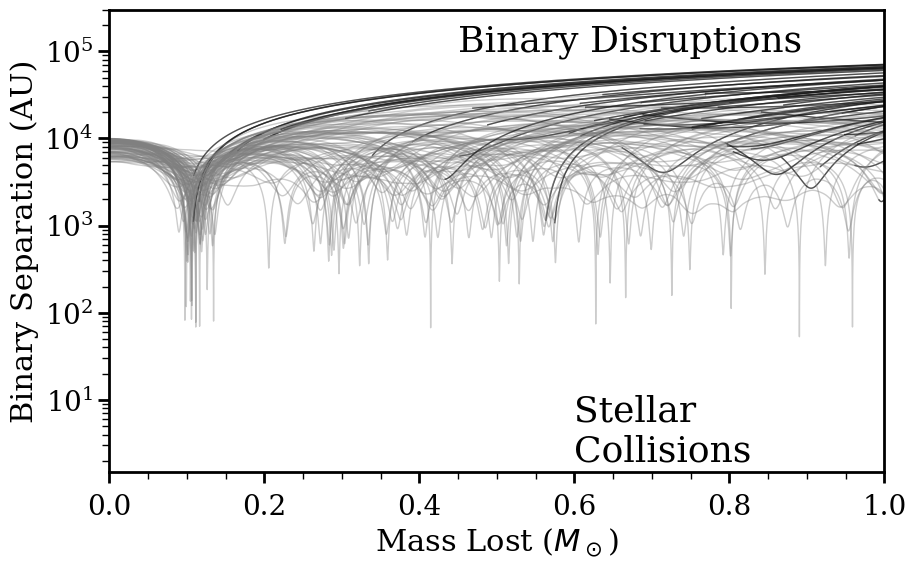}
\caption{ \label{fig:WDKickOrb} Orbital evolution of binaries undergoing episodic mass loss and stochastic kicks. Each panel shows the changing orbital separation as mass is lost for one hundred realizations. The binaries are initialized at semi-major axes of $a = 500 \, {\rm AU}$ (top), $a = 1500 \, {\rm AU}$ (middle), and $a = 5000 \, {\rm AU}$ (bottom), with a thermal distribution of initial eccentricities. Thicker black lines drifting upwards are binaries that become unbound, while lines exiting the bottom are binaries that collide or tidally circularize.}
\end{figure}

\subsection{Long-term orbital evolution}

\subsubsection{Numerical Results}
\label{sec:numerical}

Over long time scales, the stochastic series of kicks will impart a growing change to the orbital energy and angular momentum. To get a sense of the possibilities, we numerically integrate orbits undergoing random kicks, with orbital energy and angular momentum changing via equations \ref{eq:jf} and \ref{eq:ef}. We consider a fixed value $M_{\rm ej} = 10^{-4} \, M_\odot$, and we assume the kick velocity in each direction has a Gaussian distribution with standard deviation $\approx 300 \, {\rm cm/s}$. Kicks are assumed to occur on a recurrence time scale $t_{\rm kick} = t_{\rm ML}/N$, where $t_{\rm ML}$ is the mass loss time scale we assume to be $t_{\rm ML} = 10^6$ years, and $N$ is the total number of kicks. For simplicity, we consider binaries with $M_{1,i} = 1.6 \, M_\odot$, $M_2 = 1 M_\odot$, $R_1 = 300 \, R_\odot$, at a few different initial semi-major axes, and with initial eccentricities randomly sampled from a thermal distribution. We integrate the orbit for $N=10^4$ kicks such that the final WD mass is $M_{1,f} = 0.6 \, M_\odot$. We consider the binary to become unbound if its orbital energy becomes positive, and we classify it as a stellar collision if the binary separation decreases below $R_1 = 300 \, R_\odot$.

Figure \ref{fig:WDKickOrb} shows the evolving binary separation $r$ from these numerical integrations. When the initial semi-major axis is $a=500 \, {\rm AU}$, the net kick velocity $v_{\rm k,tot} \ll v_{\rm orb}$, and most of the orbits are only mildly perturbed. However, a small fraction of orbits are driven towards collisions. This outcome is more likely for orbits with higher initial eccentricities (lower initial angular momentum). 

When the initial semi-major axis is $a=1500 \, {\rm AU}$, the net kick velocity is still below $v_{\rm orb}$, so most of the orbits remain bound. However, a small fraction (about 10\%) become unbound due to the combination of mass loss and kicks. A smaller fraction (a few \%) are driven towards collision. The collision fraction is smaller than the unbound fraction because of suppression due to a full loss cone (see Section \ref{sec:collision}). When $a=5000 \, {\rm AU}$, the net kick velocity $v_{\rm k,tot} \approx v_{\rm orb}$, and so a large fraction of orbits become unbound.

\subsubsection{Analytic Analysis}
\label{sec:analytical}

To build a simple model of orbital evolution, we consider each kick to be the same magnitude, $v_k \sim (M_{\rm ej}/M_1) v_{\rm esc}$, but with a random direction. In this case, expected values of terms in equations \ref{eq:deltae} and \ref{eq:deltaj} are $\langle |\cos(\theta_k)| \rangle = 1/\sqrt{3}$, and $\langle |v_{kt}| \rangle = (M_{\rm ej}/ \sqrt{3} M_1) v_{\rm esc}$. For a high-eccentricity orbit, the time-averaged distance between the stars is $\langle r \rangle = 3 a/2$, while the average velocity is $\langle |v| \rangle = 2 v_c/\pi$ and $\langle v^2 \rangle = v_c^2$.

In this case, after each mass ejection event, the expected orbital energy change due to the first and last terms of equation \ref{eq:deltae} is
\begin{equation}
\label{eq:de}
    \bigg\langle \frac{dE}{E}\bigg\rangle \approx - \frac{M_{\rm ej}}{M_1} \left[ \frac{2 M_1 + M_2}{M_1 + M_2} + \frac{M_{\rm ej}}{M_1} \left( \frac{v_{\rm esc}}{v_c} \right)^2 \right] \, .
\end{equation}
Both terms in equation \ref{eq:de} have a fixed sign and cause the orbital energy to grow. The second term in equation \ref{eq:deltae} can cause the orbital energy to increase or decrease. The mean magnitude of the energy change due to that term is
\begin{equation}
\label{eq:de1}
\bigg\langle \left| \frac{dE}{E} \right| \bigg\rangle = \frac{4}{\pi \sqrt{3}} \frac{M_{\rm ej}}{M_1} \frac{v_{\rm esc}}{v_c} \, .
\end{equation}
Hence, that term broadens the width of the energy distribution rather than changing its mean.

Similarly, the mean angular momentum changes from the first and third terms in equation \ref{eq:deltaj} are
\begin{equation}
\label{eq:dj}
    \langle dJ \rangle  \approx - \frac{M_{\rm ej}}{M_1} \frac{M_2}{M_1 + M_2} \left[ J - \frac{1}{3} M_{\rm ej} a \, \frac{v_{\rm esc}^2}{\langle{|v_\phi|\rangle}} \right] \, ,
\end{equation}
which can cause the angular momentum to systemically increase or decrease. The mean absolute magnitude of the change in $J$ due to the second term in equation \ref{eq:deltaj} is
\begin{equation}
\label{eq:dj1}
\langle | d J | \rangle \approx \frac{\sqrt{3}}{2} M_{\rm ej} \frac{M_2}{M_1 + M_2} a \, v_{\rm esc} \, .
\end{equation}
As above, this term broadens the width of the angular momentum distribution but does not change its mean.

%Note the last terms in equations \ref{eq:de} and \ref{eq:dj} are second order in $M_{\rm ej}$ or $v_k$, and their magnitude is smaller than the terms in equations \ref{eq:de1} and \ref{eq:dj1}, but they will grow linearly with the number of kicks.

%Looking at the terms in equation \ref{eq:deltae}, we note that the first term is always negative, while the second depends on the sign of $\theta_k$ and therefore the direction of the kick. We assume each kick to be uncorrelated with a random direction, so $\cos(\theta_k)$ will have a random sign, and the second term can either increase or decrease the orbital energy. Therefore the energy will decrease monotonically due to the first term, but it will undergo a random walk due to the second term. Similarly, the first term in equation \ref{eq:deltaj} is negative while the second has a random sign and induces a random walk.

After a series of $N$ kicks, the terms in equations \ref{eq:de} and \ref{eq:dj} will grow linearly with $N$, whereas the terms in equations \ref{eq:de1} and \ref{eq:dj1} grow as $\sqrt{N}$. Equation \ref{eq:de} gives a net energy change
\begin{equation}
\label{eq:de2}
    \bigg\langle \frac{\Delta E}{E}\bigg\rangle \approx - \frac{N M_{\rm ej}}{M_1} \left[ \frac{2 M_1 + M_2}{M_1 + M_2} + \frac{M_{\rm ej}}{M_1} \left( \frac{v_{\rm esc}}{v_c} \right)^2 \right] \, ,
\end{equation}
and equation \ref{eq:de1} gives
\begin{equation}
\label{eq:de3}
\bigg\langle \left| \frac{\Delta E}{E} \right| \bigg\rangle = \sqrt{\frac{N}{3}} \frac{4}{\pi} \frac{M_{\rm ej}}{M_1}  \frac{v_{\rm esc}}{v_c} \, .
\end{equation}
Similarly, the terms in equation \ref{eq:dj} give a net angular momentum change
\begin{align}
\label{eq:dj2}
    \langle \Delta J \rangle \approx &- \frac{N M_{\rm ej}}{M_1}  \frac{M_2}{M_1 + M_2} \bigg[ J - \frac{1}{3} M_{\rm ej} a \frac{v_{\rm esc}^2}{\langle{|v_\phi|\rangle}} \bigg] \, .
\end{align}
and equation \ref{eq:dj1} gives 
\begin{equation}
\label{eq:dj3}
\langle |\Delta J | \rangle \approx \frac{\sqrt{3N}}{2} M_{\rm ej} \frac{M_2}{M_1 + M_2} a \, v_{\rm esc} \, .
\end{equation}

In other words, after $N$ kicks, the mean of the distribution in energy and angular momentum shifts by equations \ref{eq:de2} and \ref{eq:dj2} which grow linearly with $N$. The width of the distribution in energy and angular momentum is given by equations \ref{eq:de3} and \ref{eq:dj3} and grows as $\sqrt{N}$.
The first terms of equations \ref{eq:de2} and \ref{eq:dj2} have the same effect as ordinary winds, causing the orbit to gradually widen as mass is lost. The second term in equation \ref{eq:de2} is second order in $M_{\rm ej}$, but it can be comparable to the first term because $v_{\rm esc} \gg v_c$ for wide orbits. Unlike the first term, its amplitude increases as the orbit widens, and it will eventually unbind the orbit if $N$ is large enough.

%As described in Section \ref{sec:mass}, systems of interest typically have $M_1 \sim M_2 \sim M_\odot$, $R \sim 1 \, {\rm AU}$, $a \sim 10^4 \, {\rm AU}$, $v_{\rm c} \sim 0.3 \, {\rm km/s}$, and $v_{\rm esc} \sim 30 \, {\rm km/s}$. The value of $M_{\rm ej}$ is uncertain but estimated to be $M_{\rm ej} \sim 10^{-4} \, M_\odot$. To lose the entire envelope mass requires $N \sim 10^4$ mass ejection events. Hence, all the terms in equations \ref{eq:de2} and \ref{eq:dj2} have comparable magnitude and will be important for orbital evolution.

An increase of the orbital energy can contribute to orbital unbinding, while a decrease of the orbital angular momentum to nearly zero can cause stellar collisions. In binaries that are unbound, the last term of equation \ref{eq:de2} is of order unity, so $N (M_{\rm ej}/M_1)^2 (v_{\rm esc}/v_c)^2 \sim 1$. If a star loses most of its mass on the AGB, the number of ejection events is $N \sim M_1/M_{\rm ej}$. In this case, all the terms in equations \ref{eq:de2} and \ref{eq:de3} have comparable magnitude and will be important for orbital evolution.

Since the energy change in equation \ref{eq:de3} arises from steps in a random walk, the distribution of energy changes resulting from those steps converge to a Gaussian with standard deviation of 
\begin{equation}
    \sigma_E \simeq \frac{4}{\pi} \sqrt{\frac{N}{3}} \frac{M_{\rm ej}}{M_{1,f}} \frac{v_{\rm esc}}{v_{c,f}} E_f \, ,
\end{equation}
where $M_{1,f}$ is the final primary mass, $E_f = - G M_{1,f} M_2/(2 a_f)$ is the final orbital energy, $a_f$ the final semi-major axis, and $v_{c,f} = \sqrt{G (M_{1,f} + M_2)/a_f}$. We find that using the final mass and semi-major axis (rather than initial properties) provides a better match with numerical results. The final semi-major axis due to orbital expansion from the mass loss is 
\begin{equation}
a_f = a_i \frac{M_{1,i}+M_2}{M_{1,f}+M_2} \, .
\end{equation}

The expected change in energy resulting from the last term in equation \ref{eq:de2} is
\begin{equation}
    E_2 = \frac{N}{2} \frac{M_{1,f} M_2}{M_{1,f} + M_2} \left(\frac{M_{\rm ej}}{M_{1,f}} v_{\rm esc} \right)^2 \, ,
\end{equation}
and we used $E = - G M_{1,f} M_2/(2a_f)$ and $v_c^2 = G (M_{1,f} + M_2)/a_f$. Hence the expected distribution of energy after $N$ kicks is a Gaussian centered at energy $E = E_f + E_2$,
\begin{equation}
    p(E) = \frac{1}{\sqrt{2 \pi \sigma_E^2}} {\rm exp} \left(\frac{-(E-E_f-E_2)^2}{2 \sigma_E^2}\right) \, .
\end{equation}

The fraction of systems that become unbound are those that have $E > 0$ after a series of $N$ kicks. This fraction is
\begin{align}
\label{eq:fdis}
    f_{\rm un,0} &= \int^\infty_0 p(E) dE \nonumber \\
    &= \frac{1}{2} \left[ 1 - {\rm erf} \left(\frac{- E_f - E_2}{\sqrt{2} \sigma_E}\right) \right] \, ,
\end{align}
where ${\rm erf}$ is an error function.

\begin{figure}
\includegraphics[scale=0.37]{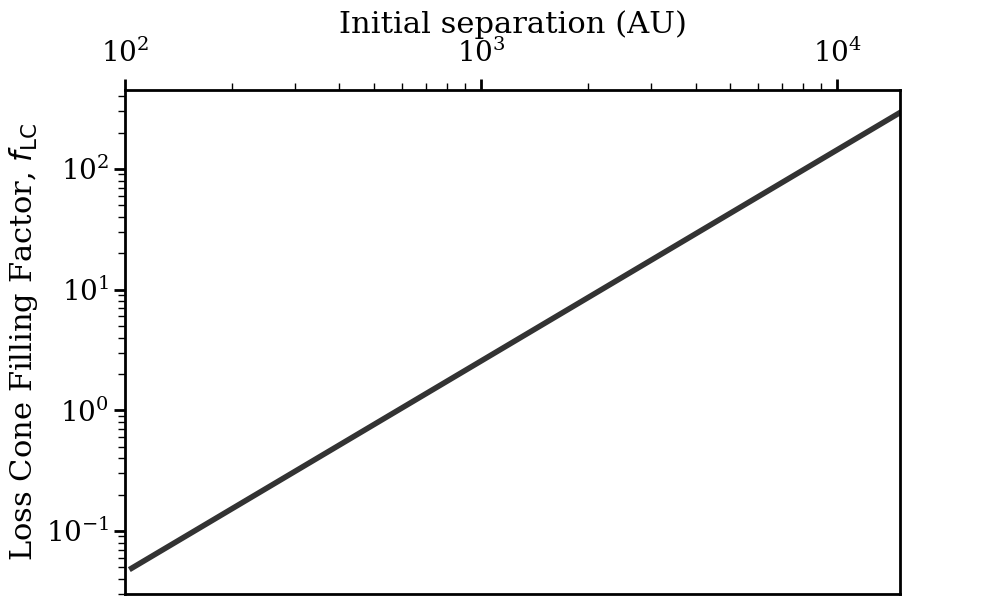}
\includegraphics[scale=0.37]{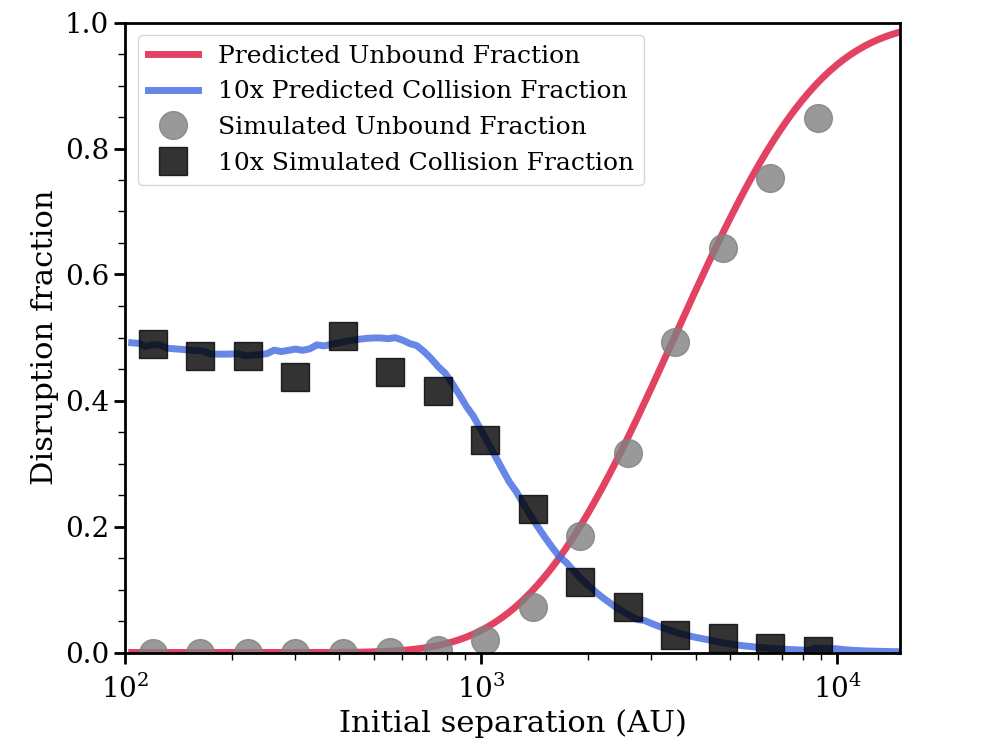}
\caption{ \label{fig:WDKickOrbFrac} Top: loss cone filling factor $f_{\rm LC}$ as a function of initial semi-major axis, given individual kick amplitudes of 5 m/s and a $1.6 \, M_\odot$ star with a $1 \, M_\odot$ companion. Bottom: binary disruption fractions vs. initial semi-major axis for the same binary parameters. Gray points are unbound systems from orbital evolution simulations, while the red line is an analytic estimate. Black points are systems that experienced a collision, compared to the analytical estimate shown by the blue line.}
\end{figure}

Figure \ref{fig:WDKickOrbFrac} compares the prediction of equation \ref{eq:fdis} to the numerical integrations described in Section \ref{sec:numerical}, for different initial semi-major axes $a_i$. We perform 4000 integrations at each value of $a_i$ and compute the fraction of systems that become unbound. We can see that the predicted unbound fraction matches the numerical results remarkably well.

Mass ejection events from real stars will vary in both direction and magnitude. While our analytics consider constant ejecta masses and kick magnitudes, our numerical results randomly draw kick speeds from a normal distribution of standard deviation $v_{\rm kick} = M_{\rm ej}/(\sqrt{3} M_1 v_{\rm esc})$ in each direction. We have found that allowing for varying $M_{\rm ej}$ or fixed $v_{\rm kick}$ does not significantly change the numerical results, suggesting our analytic calculations apply well to real stars whose momenta vary in both magnitude and direction.

\subsubsection{Stellar Collisions}
\label{sec:collision}

We can estimate the number of stellar collisions in a similar fashion, realizing that the angular momentum also undergoes a random walk. In this case, however, the random walk occurs in two dimensions. Collisions occur when the angular momentum decreases below a minimum value
\begin{equation}
    \label{eq:jmin}
    J_{\rm min} = M_1 M_2 \sqrt{2 G R_1/(M_1+M_2)}
\end{equation}
which applies to an orbit with $e \approx 1$ and periastron radius $r_{\rm peri} = a (1-e) = R_1$, where $R_1$ is the radius of the AGB star. This corresponds to the orbit entering the ``loss cone", very similar to tidal disruption events in galactic centers, investigated in many prior works (e.g., \citealt{cohn:78,bar-or:16,stone:20}). 

%From equation \ref{eq:dj2}, the expected distribution of angular momenta after $N$ steps is 
%\begin{equation}
%    p(J) = \frac{1}{\sqrt{2 \pi \sigma_J^2}} {\rm exp} \left(\frac{-(J-J_i)^2}{2 \sigma_J^2}\right) \, ,
%\end{equation}
%where
%\begin{equation}
%\label{eq:sigmaJ}
%    \sigma_J = \frac{\sqrt{3 N}}{2}  \frac{M_{\rm ej}}{M_1} \frac{M_1 M_2}{M_1 + M_2} a \, v_{\rm esc} \, .
%\end{equation}
%As above, to account for ongoing mass loss we replace $M_1$ with $M_{1,ef}$ and $a$ with $a_{\rm ef}$, and we use $J_i = M_{1,{\rm ef}} M_2 \sqrt{G a_{\rm ef} (1-e_i^2)/(M_{1,{\rm ef}}+M_2)}$, where $e_i$ is the initial orbital eccentricity.

We define the standard deviation of the angular momentum change similarly to that of energy, using equation \ref{eq:dj3},
\begin{align}
\label{eq:sigmaJ}
    \sigma_J &= \frac{\sqrt{3 N}}{2}  \frac{M_{\rm ej}}{M_{1,f}} \frac{M_{1,f} M_2}{M_{1,f} + M_2} a_f \, v_{\rm esc} \nonumber \\
    &= \sqrt{N} \Delta J_k \, ,
\end{align}
and $\Delta J_k$ is the typical angular momentum change from a single kick,
\begin{equation}
\label{eq:deltaJ}
\Delta J_k = \frac{\sqrt{3}}{2}  \frac{M_{\rm ej}}{M_{1,f}} \frac{M_{1,f} M_2}{M_{1,f} + M_2} a_f \, v_{\rm esc} \, .
\end{equation}
During each orbit, the typical change in angular momentum is
\begin{equation}
    \Delta J_{\rm orb} \simeq \sqrt{N_{\rm k,orb}} \Delta J_{\rm k}
\end{equation}
where $N_{\rm k,orb} = N P_{\rm orb}/t_{\rm ML}$ is the number of kicks per orbit, $P_{\rm orb}$ is the orbital period, and $t_{\rm ML}$ is the mass loss time scale.

One may naively guess that collisions become likely when the expected change in angular momentum, $\sigma_J$, becomes larger than the initial orbital angular momentum. This yields the wrong answer in much of the parameter space because the angular momentum change per orbit, $\Delta J_{\rm orb}$, can be much larger than $J_{\rm min}$. This is known as the full loss cone regime, where the probability of a collision decreases as $(\Delta J_{\rm orb}/J_{\rm min})^{-2}$. We define the loss cone filling factor 
\begin{align}
    f_{\rm LC} &= \frac{\Delta J_{\rm orb}}{J_{\rm min}} \nonumber \\
    &\sim 10  \, \left(\frac{t_{\rm ML}}{10^{6} \, {\rm yr}} \right)^{\!-1/2} \left(\frac{M_{\rm ej}}{10^{-4} M_\odot}\right)^{\!1/2}  \left(\frac{a}{1000 \, {\rm AU}}\right)^{\!7/4}\, .
\end{align}
for a fiducial system with $M_1 = 1.6 \, M_\odot$, $M_2 = 1 \, M_\odot$, $R_1 = 300 R_\odot$. We thus see that most strongly perturbed binaries with $a \gtrsim 10^3 \, {\rm AU}$ will be in the full loss cone regime, while tighter binaries can be in the empty loss cone regime. Figure \ref{fig:WDKickDisruptionFrac} plots the loss cone filling factor for our fiducial AGB parameters, showing that $f_{\rm LC} > 1$ for semi-major axes $a \gtrsim 500 \, {\rm AU}$.

Following the same logic as equation \ref{eq:fdis}, but estimating the fraction of orbits that diffuse to $J=0$ and accounting for suppression of collisions in the full loss cone regime, a rudimentary estimate for the collision fraction is
\begin{equation}
\label{eq:fcol}
    f_{\rm col}(e_i) = \frac{1}{2} \left[ 1 - {\rm erf} \left(\frac{J_i(e_i)}{\sqrt{2} \sigma_J}\right) \right] \frac{1}{(1 + f_{\rm LC})^2} \, ,
\end{equation}
where $e_i$ is the initial orbital eccentricity. We find this estimate is okay, roughly a factor of $\sim$2 below that of our numerical integrations shown in Figure \ref{fig:WDKickOrbFrac}.

A more accurate estimate for the collision probablilty can be achieved using calculations developed for galactic centers, which solve the steady state Fokker-Planck distribution of stellar orbits including a flux of objects into the loss cone. At a given semi-major axis, \cite{stone:20} find a tidal disruption rate
\begin{equation}
\label{eq:stone1}
    \frac{dN_{\rm col}}{dt} = \frac{D_J/J_c^2}{(\alpha -1)(1-R_c) + \ln(1/R_c)}  \, .
\end{equation}
Here, $J_c = M_1 M_2 \sqrt{G a/(M_1 + M_2)}$ is the angular momentum of a circular orbit at semi-major axis $a$, $D_J$ is the angular momentum diffusion coefficient, $R_c = (J_{\rm min}/J_c)^2$, and $\alpha$ accounts for full loss cone effects, with $\alpha \approx (f_{\rm LC}^2 + f_{\rm LC}^4)^{1/4}$.

The diffusion coefficient in our case is determined by kicks. For a two-dimensional random walk, the relation between standard deviation and diffusion coefficient is
\begin{equation}
     D_J = \frac{\sigma_J^2}{4 \tau} \, ,
\end{equation}
where $\tau$ is the duration of the process. Over a time $\tau$, from equation \ref{eq:stone1} the expected collision fraction at a given semi-major axis would be
\begin{equation}
\label{eq:fcol1}
    f_{\rm col} = \tau \frac{dN_{\rm col}}{dt} = \frac{\sigma_J^2 / 4 J_c^2}{(\alpha -1)(1-R_c) + \ln(1/R_c)} \, .
\end{equation}
We find this equation substantially underestimates the rate of collisions compared to our numerical results, because it applies to a steady state where orbits are continually diffusing to small values of $J$. In our case, individual systems do not have enough time to reach a steady state distribution, and colliding systems are those that begin with small angular momentum.

We can much more accurately match the numerical results if we apply equation \ref{eq:fcol1} to systems using their initial conditions. In this case, the collision fraction at a given initial orbital angular momentum $J_i = J_c \sqrt{1 - e_i^2}$ is
\begin{equation}
    f_{\rm col}(e_i) = \frac{\sigma_J^2 / 4 J_i^2}{(\alpha -1)(1-R_i) + \ln(1/R_i)} \, ,
\end{equation}
where $R_i = (J_{\rm min}/J_i)^2$. The total collision fraction $f_{\rm col}$ at a given semi-major axis, accounting for a distribution of eccentricities $e_i$, is then
\begin{equation}
\label{eq:fcoltot}
    f_{\rm col} = \int^1_0 f_{\rm col}(e_i) p(e_i) d e_i \, ,
\end{equation}
and again we assume an initially thermal eccentricity distribution.

Figure \ref{fig:WDKickOrbFrac} compares this prediction with our simulations, showing good agreement. The collision fraction is small, always less than 10\% for these parameters, and it peaks at small semi-major axes. The ratio $\sigma_J/J_i$ scales as $a^{1/2}$ so one might expect more collisions at larger semi-major axes. However, the loss cone filling factor becomes large at large semi-major axes, suppressing collisions.

When the loss cone filling factor is small, the periastron separation shrinks gradually, so tides in the AGB star may be able to circularize the orbit before a collision occurs. The true number of collisions will thus be smaller than shown in Figure \ref{fig:WDKickOrbFrac}, especially at smaller semi-major axes. Conversely, our collision criterion $r_{\rm peri} < R_1$ likely underestimates the rates of disruptions, because orbits with $r_{\rm peri} \lesssim 3 R_1$ can lead to tidal disruption or tidal circularization. More detailed work incorporating tidal effects into the orbital evolution will be needed to understand the relative rates of tidal circularization, tidal disruption, and true stellar collisions.

%The overprediction at large semi-major axis likely results from the fact that most orbits can be unbound before they are driven to collision, which is not accounted for in the model.

%We find the estimate of equation \ref{eq:fcol} typically underpredicts the disruption fraction at semi-major axes of $\sim \! 5000 \, {\rm AU}$. We suspect this occurs because the simple model above does not account for the changing effects of kicks as the orbit becomes eccentric. As can be seen in equation \ref{eq:deltaj}, the relative kick-induced changes in angular momentum increase as the tangential orbital velocity, $v_t$, decreases. Since the typical value of $v_t$ decreases as an orbit becomes more eccentric at the same semi-major axis, our model likely underpredicts the diffusion rate to $J=0$. Our analytical estimate is therefore likely to be a lower limit, and this issue should be studied more carefully in the future.

\vspace{-10pt}
\subsection{Binary Disruption Rates}

\begin{figure}
\includegraphics[scale=0.32]{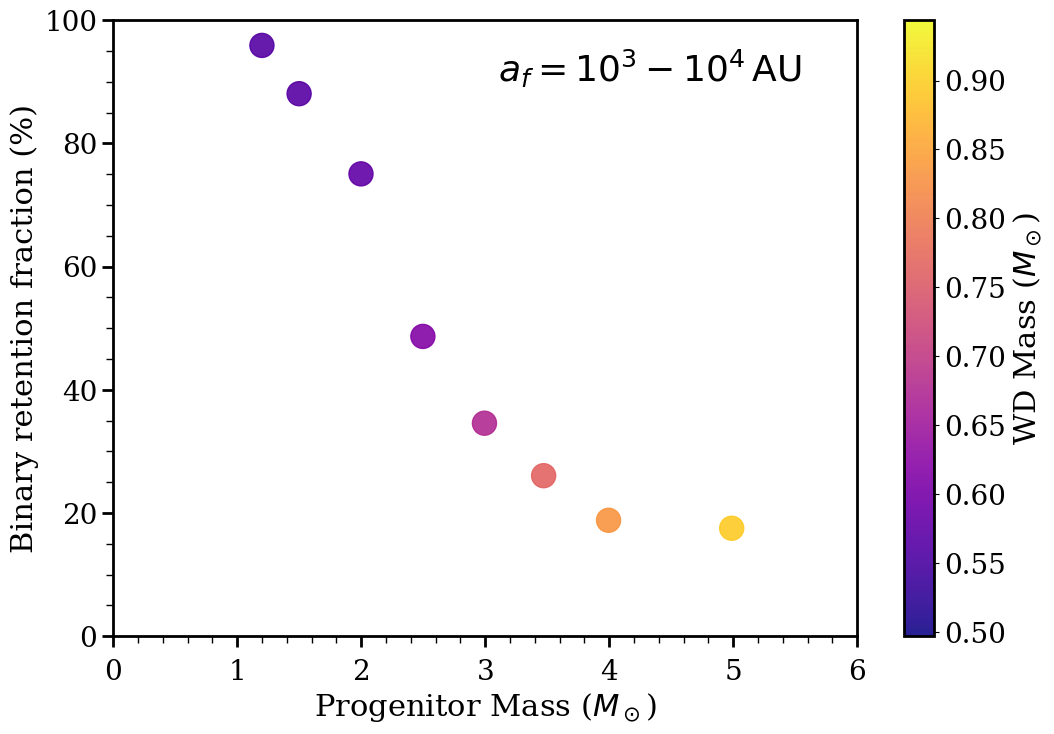}
\includegraphics[scale=0.32]{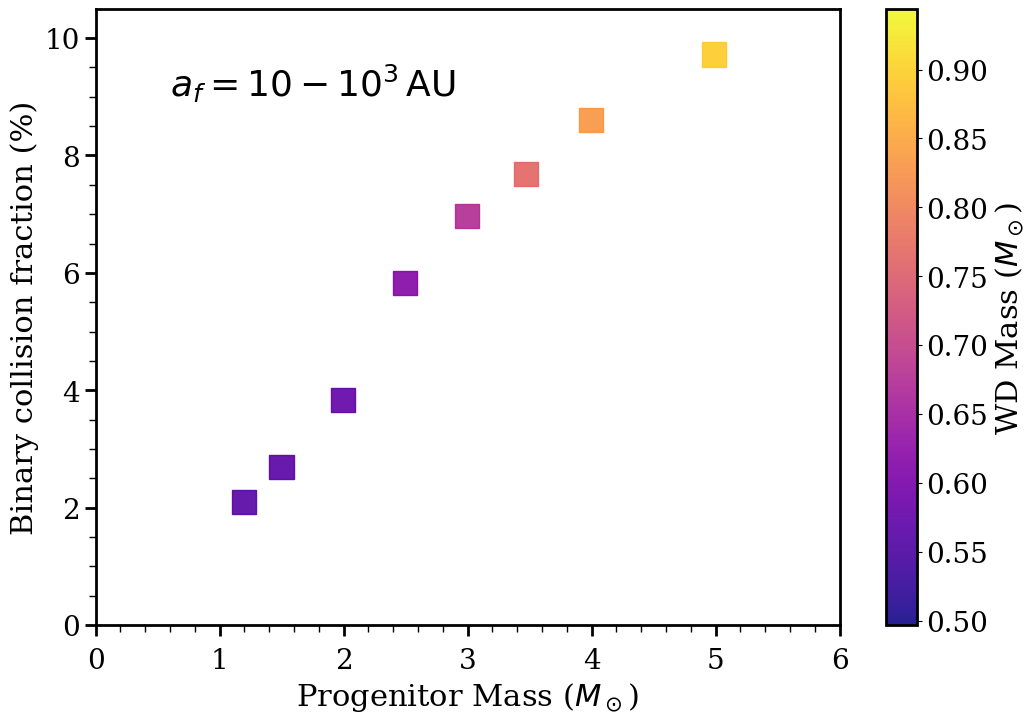}
\caption{ \label{fig:WDKickDisruptionFrac} Top: the predicted fraction of surviving WD-MS binaries with final separations in the range $a_f = 10^3 - 10^4 \, {\rm AU}$, as a function of progenitor mass, using net kick amplitudes shown in Figure \ref{fig:WDKicks}. These retention fractions are similar to those measured by \citealt{hwang:25}. Bottom: fraction of binaries that collide or tidally circularize as a function of progenitor mass for binaries with final separations in the range $a_f = 10 - 10^3 \, {\rm AU}$. }
\end{figure}

Based on the analytical results of Section \ref{sec:analytical} and \ref{sec:collision}, we can compute the expected rate at which binary orbits are unbound or driven towards collision. To do this for realistic stars, we use the net kick $v_{k,{\rm tot}}$ from equation \ref{eq:vktotint} to compute the change in energy via $E_2 = (1/2) M_{1,f} M_2 v_{k,{\rm tot}}^2/(M_{1,f}+ M_2)$, and the standard deviation $\sigma_E = (4/ \pi \sqrt{3}) (v_{k,{\rm tot}}/v_{c,f}) E_f$. Similarly, the angular momentum standard deviation is $\sigma_J = (\sqrt{3}/2) M_{1,f} M_2 a_f \, v_{k,{\rm tot}}/(M_{1,f}+M_2)$. As in Figure \ref{fig:WDKicks}, we use $f_{\rm ej} = 0.5$. The initial semi-major axis distribution is assumed to be the log-normal distribution of \cite{duquennoy:91}, and we set $M_2 = M_{1,i}/2$.

%The final semi-major axis (accounting for mass loss but not kicks) of each binary is defined to be $a_f = a_i (M_{1,i} + M_2)/(M_{1,f}+M_2)$, where $M_{1,i}$ and $M_{1,f}$ are the ZAMS mass and WD mass of the primary.

For each stellar mass, we then compute the analytical integrals for unbound orbits and stellar collisions from equations \ref{eq:fdis} and \ref{eq:fcoltot}, integrating over a thermal initial eccentricity distribution for the latter. To compare with measurements from \cite{hwang:25} (see their Figure 4), we compute the fraction of surviving binaries with final semi-major axes $a_f = 10^3 - 10^4 \, {\rm AU}$.

Figure \ref{fig:WDKickDisruptionFrac} shows our results. The  fraction of surviving binaries decreases strongly with progenitor mass, from nearly 100\% at $1 \, M_\odot$ to less than 20\% at $5 \, M_\odot$. In our models, high-mass WDs are more often unbound due to their larger kicks shown in Figure \ref{fig:WDKicks}. The collision fraction also increases with mass, from roughly 2\% for $1 M_\odot$ AGB stars to nearly 10\% for high-mass AGB stars.

%\cite{hwang:25} find that the fraction of disrupted systems increases from near zero for low-mass WDs to $\sim$90\% for high-mass WDs, very similar to our results.

\section{Discussion and Conclusions}

Our model makes several testable predictions that may discriminate it from other possibilities. We predict net kick amplitudes of $\sim$1 km/s, but this value is quite uncertain because it scales with $\sqrt{M_{\rm ej}}$, and the value of $M_{\rm ej}$ is uncertain at the order of magnitude level. A more robust prediction is that the kick is strongly correlated with WD mass, as shown in Figure \ref{fig:WDKicks}. We therefore predict that the fraction of WDs in wide binaries falls off more sharply for high-mass WDs compared to low-mass WDs, as shown in Figure \ref{fig:WDKickDisruptionFrac}. These results are quite similar to the disruption fractions found by \cite{hwang:25} from \textit{Gaia} data, demonstrating the viability of this model. We achieve good agreement with the measurements by choosing $f_{\rm ej} = 0.5$ (equation \ref{eq:Mej}), which we deem as a success for our model, because we expect $f_{\rm ej}$ to be of order unity. The predicted fraction of disrupted binaries also increases sharply with semi-major axis (Figure \ref{fig:WDKickOrbFrac}) in a manner similar to that observed by \cite{elbadry:18} and \cite{hwang:25}. More robust predictions will require a more realistic distribution of binary semi-major axes and companion masses that vary with primary mass.

%For WDs of mass $M\approx 0.6 \, M_\odot$, our models predict a steadily increasing fraction of disrupted binaries with semi-major axis, with most binaries becoming unbound at $a \gtrsim 4 \times 10^3 \, {\rm AU}$. This number is uncertain because it depends on the uncertain value of $M_{\rm ej}$.
If our fiducial estimate of $M_{\rm ej} \sim 10^{-4} \, M_\odot$ is too small/large, the number or disrupted systems will increase/decrease. For smaller/larger $M_{\rm ej}$, the disrupted systems will shift to larger/smaller separations, but the shape of the distribution should remain similar. We predict disrupted binaries will occur at smaller semi-major axes for high-mass WDs, which can likely be tested with larger samples of wide WD binary systems from Gaia DR4. In close binaries with $a \lesssim 3 \times 10^2 \, {\rm AU}$, WD kicks could increase the orbital eccentricity, potentially accounting for the eccentricities of some barium stars \citep{izzard:10}. However, kick amplitudes of $\sim$1 km/s are not enough to account for the observed eccentricities of $\sim$1 AU WD binaries detected with Gaia \citep{chawla:25}, and eccentricity excitation via mass transfer \citep{parkosidis:26} or common envelope events \citep{sandquist:98} may be necessary to explain those systems.

WD kicks can be larger than the escape velocities from open clusters, so we predict that open clusters with escape velocities less than a few km/s will have a deficit of WDs. Indeed, there is already substantial evidence for missing WDs in open clusters \citep{weidemann:77,richer:21,grondin:24,yan:26}, which can be explained if they receive small kicks of order a couple km/s \citep{fellhauer:03}. Observations of globular clusters also indicate the WD population is more extended than expected \citep{davis:08}, which can again be explained by WD kicks of order a couple km/s \citep{heyl:07a,fregeau:09}. We predict that cluster WD deficits will be stronger for high-mass WDs, which may be consistent with the observed lack of high-mass WDs in open clusters \citep{richer:21}. High-mass WDs have recently been identified escaping clusters with velocities of a few km/s \citep{heyl:22,miller:22,miller:23}, in line with our expectations. We predict high-mass WDs will be escaping at systematically larger velocities compared to low-mass WDs.

A few recent studies (\citealt{hwang:25,oconnor:26}, see also \citealt{heyl:07b}) have investigated WD kicks from asymmetric mass loss. Those models are fundamentally different because they consider mass loss with a persistent directionality, producing a steady rocket effect, rather than a series of randomly oriented small kicks (i.e., stochastic rockets).
\footnote{To their credit, \cite{oconnor:26} does mention that kicks may occur via mass ejection events in random directions rather than steady rockets.} 
The steady rocket effect can unbind binaries only if the mass loss time scale is shorter than an orbital period, which requires $a \gtrsim 10^4 \, {\rm AU}$ for a mass loss time of $10^6 \, {\rm yr}$. At closer separations, the rocket effect can instead pump the eccentricity of a binary, which could cause collisions or tidal circularization when the periastron separation approaches the radius of the AGB star.

\cite{oconnor:26} found that tidal circularization or collisions could occur in up to $\sim$40\% of WD binaries with $a < 10^3 \, {\rm AU}$, creating a paucity of WD binaries with very large eccentricities. In contrast, our mechanism predicts a small effect on binaries with $a < 10^3 \, {\rm AU}$ for our fiducial value of $M_{\rm ej} = 10^{-4} \, M_\odot$, although it could cause up to $\sim \! 10\%$ of such binaries to collide or tidally circularize. A problem with the rocket model of \cite{oconnor:26} is that it preferentially reduces the number of short-period WD binaries ($a \lesssim 300 \, {\rm AU}$), whereas \cite{elbadry:18} and \cite{hwang:25} find that it is long-period binaries ($a \gtrsim 3 \times 10^3 \, {\rm AU}$) that are preferentially disrupted.

\cite{hwang:25} also found that a simple rocket model could not reproduce the observed higher disruption fraction of long-period and high-mass WD binaries. To produce this trend, they appealed to a mass-dependent time scale of AGB mass loss, decreasing from $\sim \! 2 \times 10^6$ years for low-mass WD progenitors to less than $\sim \! 10^4$ years for high-mass WD progenitors. In their model, the binaries become unbound because the mass loss time scale becomes shorter than an orbital period. While this is certainly a possibility, our model naturally accounts for the increasing binary disruption fraction with mass (Figure \ref{fig:WDKickDisruptionFrac}) without introducing additional adjustable parameters, making it an appealing solution.

A limitation of our calculations is that we have assumed the kicks occur at random orbital phases. This is likely a good assumption when the mass loss time scale is longer than an orbital period, but it will fail for $a \gtrsim 10^4 \, {\rm AU}$ for AGB mass loss rates of $10^{-6} \, M_\odot/{\rm yr}$, or $a \gtrsim 500 \, {\rm AU}$ for mass loss rates of $10^{-4} \, M_\odot/{\rm yr}$. When the mass loss time scale is shorter than an orbital period, all the episodic mass loss events will occur at nearly the same orbital phase, approaching the limit of a single impulsive kick. The transition between these regimes should be investigated in future work.

%Our fiducial model with $M_{\rm ej} \sim 10^{-4} M_\odot$ naturally predicts binary disruption at $a \gtrsim 3 \times 10^3 \, {\rm AU}$ in rough agreement with \cite{elbadry:18}. While our models do not include a persistent source of asymmetry (e.g., due to stellar rotation or a magnetic field), it remains possible that such asymmetry does occur, in which case both a rocket effect and kick effects would occur simultaneously.

%Another observational diagnostic of mass loss from AGB stars are stellar dimming events like the one observed for Betelgeuse in 2020. The dimming is believed to result from dust formation in the outflow following an episodic mass ejection event \citep{dupree:22}. In Betelgeuse, the dimming events appear to occur roughly once every $\sim$100 years, indicating they eject $\sim \! 10^{-4} \, M_\odot$ at a time in order to account for Betelgeuse's inferred mass loss rate of $\sim \! 2 \times 10^{-6} \, M_\odot$/yr \citep{decin:12}. These dimming events have also been observed for AGB stars (e.g., \citealt{bedding:02}), which can lose mass at similar rates to RSG stars.

In our orbital models, a ``collision" occurs when the periastron radius becomes smaller than an extended AGB star with $R \sim 300 \, R_\odot$. 
%If the periastron separation shrinks gradually, tidal circularization may instead occur. In Section \ref{sec:collision}, we estimated a typical loss cone filling fraction of $f_{\rm LC} \sim 1$, meaning that the periastron separation can change by order unity in subsequent orbits. This suggests that both tidal circularization and direct collisions can occur, with collisions more likely in systems with larger semi-major axes. 
When they occur, such collisions might be better described as AGB tidal disruption events, and we expect such events will frequently result in eccentric common-envelope events, which can produce post-common envelope binaries with larger separations and higher eccentricities than those produced by circular binaries \citep{glanz:21}. Such events may contribute to the emerging population of eccentric post-CE binaries found with \textit{Gaia} data \citep{yamaguchi:24}. Common envelopes likely produce some sort of luminous red nova \citep{ivanova:13}, and events arising from the more violent eccentric cases may be systematically brighter \citep{soker:06}.

The event rates of such high-eccentricity common envelope events is likely substantial. From Figure \ref{fig:WDKickDisruptionFrac}, we find that roughly $\sim 2-10 \%$ of binaries in the range $a= 10-10^3 \, {\rm AU}$ result in collisions, and roughly 70\% of binaries are born in this range based on \cite{duquennoy:91}. Since more than $\sim$50\% of WD-producing stars originate in binaries, this implies that the collision rate is of order a few percent the WD birth rate. Given an expected common envelope rate of $\sim$5\% of the WD birth rate \citep{politano:10,ivanova:13}, high-eccentricity collisions with AGB stars may contribute to a significant fraction of common envelope events. However, we again emphasize that many of these ``collisions" may actually result in tidal circularization, especially for systems beginning at smaller semi-major axes.

WD kicks will affect any orbiting planets or planetesimals. Kicks of $\sim$1 km/s will unbind most comets originating from Oort clouds \citep{oconnor:23} and may help account for the paucity of WDs accreting water-rich material. Such kicks will not unbind planets with $a \lesssim 10^2 \, {\rm AU}$, so they are unlikely to account for the rarity of polluted high-mass WDs \citep{ouldrouis:24,cunningham:25}. However, the kicks will increase planetary orbital eccentricities, which may seed subsequent orbital instabilities (\citealt{stephan:26}, see \citealt{veras:24} for a review on polluted WDs).

We have not investigated effects of episodic mass loss in triple systems. \cite{shariat:23} modeled effects of single instantaneous WD kicks on triple systems, finding that kicks of $\sim$1 km/s unbind outer tertiaries with $a \gtrsim 3000 \, {\rm AU}$, similar to the effects on binary systems. They also showed that the observed distribution of triple separations from \textit{Gaia} data provides strong evidence for kicks of this magnitude. Episodic mass loss may introduce new pathways for triple evolution just as it does for binaries, and this should be investigated in future work.

\section*{Acknowledgments}

I thank Re'em Sari, Jing-Ze Ma, Kareem El-Badry, and Hagai Perets for useful discussions. This research was supported in part by grant NSF PHY-2309135 to the Kavli Institute for Theoretical Physics (KITP).

\section*{Data Availability}

The orbital integration codes and plotting scripts to make Figures 2-5 are available upon request.

\bibliography{bib}

\end{document}